\documentclass[usenatbib, useAMS]{mn2e}
\input epsf
\newcommand{\ion}[2]{#1\,{\sc #2}}
\newcommand{\object}{}
\newcommand{\litanf}{\begin{list}{}{\leftmargin=1.0cm \rightmargin=0cm
\itemindent=-1.0cm \parsep=0cm \itemsep=0cm }}
\newcommand{\litend}{\end{list}}
\newcommand{\loggw}[1]{\mbox{$\log g\hspace{-0.5mm} =\hspace{-0.5mm}  #1$}}
\newcommand{\Teff}{\mbox{$T_\mathrm{eff}$}}   
\newcommand{\Teffw}[1]{\mbox{$\Teff\hspace{-0.5mm} =\hspace{-0.5mm} #1 \mathrm{,000~K}$}}

\newcommand{\lapprox}{\lower0.8ex\hbox{$\buildrel <\over\sim$}}
\newcommand{\gapprox}{\lower0.8ex\hbox{$\buildrel >\over\sim$}}
\usepackage{epsfig}
\usepackage{epsf}
\usepackage{times}
\input epsf
\def \ul #1:#2:{$^{+#1}_{-#2}$}

\title[3-D photoionization modelling of the hydrogen-deficient knots in the planetary nebula Abell~30]{Three-dimensional photoionization modelling of the hydrogen-deficient knots in the planetary nebula Abell~30}
\author[Ercolano et al. ] {B. Ercolano$^1$, M. J. Barlow$^1$, P. J. Storey$^1$, X.-W. Liu$^2$, T. Rauch$^{3,4}$, K. Werner$^4$\\
$^1$University College London, Gower Street, London WC1E 6BT, UK\\
$^2$Department of Astronomy, Peking University, Beijing, 100871, China\\
$^3$Dr.-Remeis-Sternwarte, Sternwartstra\ss e 7, D-96049 Bamberg, Germany\\
$^4$Institut f\"ur Astronomie und Astrophysik, Abt. Astronomie, Sand 1, Universit\"at T\"ubingen, 72076 T\"ubingen, Germany\\}

\date{Received:}
\begin{document}
\maketitle
\begin{abstract}
\noindent

We have constructed a photoionization model, using the 3-D Monte Carlo code \mbox{\sc mocassin} for one of the hydrogen-deficient knots (J3) of the born-again planetary nebula Abell~30. The model consists of spherical knots, comprising a cold, dense, hydrogen-deficient core with very high metal abundances. The inner core, occupying 9.1\% of the total volume of the knot, is surrounded by a less dense hydrogen-deficient and metal-enriched gas envelope, with less extreme abundances. The envelope of the knot might have been formed by the mixing of the knot material with the surrounding nebular gas.

This bi-chemistry, bi-density model did not produce enough heating to match the fluxes of the collisionally excited emission lines (CELs) and of the optical recombination lines (ORLs) observed in the spectrum of the knot. We therefore included heating by photoelectric emission from dust grains in the thermal equilibrium calculations, and found that dust-to-gas ratios of 0.077 and 0.107 by mass for the central core and the envelope of the knot, respectively, are sufficient to fit the spectrum. Surprisingly, photoelectric emission from grains is the dominant source of heating in the hot envelope of the knot, while heating by photoionization of helium and heavy elements dominates in the cold core. 

We obtain a good fit with the observations for most of the significant emission lines treated in our model. The two major discrepancies occurred for the [O~{\sc ii}]~3727,29~{\AA} doublet and the [N~{\sc ii}]~6548, 6584~{\AA} lines, which are severely underestimated in our model. Recombination contributions could be significant and we included them for the O~{\sc ii} transitions. However, this was not sufficient to resolve the discrepancy, due to the high collisional de-excitation rates in the dense core, where most of the recombination lines would be produced. This possibly highlights a weakness using a discontinuous density distribution like ours, where in reality one might expect an intermediate phase to exist. 

 The chemical abundances inferred from our modelling of the central core region and of the envelope of the knot are, at least qualitatively, in agreement with the abundances derived by the empirical analysis of \citet{wesson03}, although the discrepancies between the core and the envelope abundances that we find are less dramatic than those implied by the ORL and CEL empirical analysis. Our models also indicate, in agreement with the empirical analysis of Wesson et al. (2003), that the C/O ratio in the two regions of the knot is less than unity, contrary to theoretical predictions for born-again nebulae \citep[e.g.][]{iben83}.

\end{abstract}


\section{Introduction}
\label{sec:abell30intro}

The large angular diameter planetary nebula Abell~30 belongs to the very peculiar class of planetary nebulae with hydrogen-rich envelopes surrounding hydrogen-deficient inner knots \citep{jacoby79} -- other examples include Abell~78 \citep{jacoby79} and Abell~58 \citep{pollacco92}. The outer envelope of Abell~30 has a radius of 64$''$ and an almost perfectly round shape, showing a high excitation and a very low surface brightness. Given the size of the outer envelope and the fact that it appears to be very diffuse, Abell~30 is believed to be quite old \citep[$\sim$10,000 yr, if a distance of 1.3~kpc is assumed;][]{harrington84}. Its hydrogen-deficient central star, however, still appears to be quite bright (V=14.3). These characteristics, and, in particular,  the presence of four symmetrically placed hydrogen-deficient inner knots led \citet{ibenkaler83} to the classification of this object as a planetary nebula with a {\it born-again} nucleus. The central knots were identified simultaneously by \citet{jacoby79} and \citet{hazard80}; according to the {\it born-again} scenario, these knots consist of material whose abundance distribution has been altered by complete hydrogen burning and later ejected when the star returned to a brief red-giant phase after a final He-shell flash. The mechanism by which these knots would have been emitted is not known. The strongly axially symmetric structure of Abell~30 led \citet{livio95} to suggest that tidal interaction of the central star with a companion may be a plausible mechanism. Following Jacoby's discovery, the knots were named J1 to J4, clockwise, starting from the southernmost one. 

Since the discovery of the hydrogen-deficient knots, Abell~30 has been widely studied. \citet{jacoby83} obtained optical spectra and derived abundances for knots J3 and J4. \citet{harrington84} obtained IUE spectra of the nebulosity near the nucleus, in proximity to one of the knots, and carried out a study of the chemical abundances from the observed emission lines. More recently, \citet{guerrero96} published a study of the chemical abundances of hydrogen-poor planetary nebulae, including Abell~30, based on long-slit data obtained in 1986 at the 3.6\,m telescope at ESO, La Silla Observatory (Chile). \citet{wesson03} have analyzed new observational data, including optical spectra of the polar knots J1 and J3 and of the central star of Abell~30, obtained in 2000 with the ISIS spectrograph mounted on the 4.2\,m WHT at the Observatorio del Roque de los Muchachos, on La Palma, Spain, as well as archive ultraviolet spectra of knots J3 and J4 obtained in 1994 with the Faint Object Spectrograph aboard the Hubble Space Telescope.  They found large discrepancies between the ionic abundances measured from optical recombination lines (ORLs) and those measured from collisionally excited lines (CELs). This had already been noticed by \citet{jacoby83} who found a very high carbon abundance (almost half that of hydrogen) from ORLs, far higher than that derived from CELs. Subsequently, \citet{harrington84} proposed that this discrepancy could be explained if very large temperature variations were present within the knots of Abell~30. A high concentration of heavy elements in the core of the knots could create a very cool ($\approx$1000~K) but still ionized region emitting the observed ORLs. This could be surrounded by a hotter region consisting of material with more typical nebular abundances emitting the observed CELs. The analysis of \citet{wesson03} is in agreement with this interpretation. 

In Section~2 of this paper we give a description of the photoionization modelling, together with details of the parameters used. A strategy which was used in order to estimate the effects of photoelectric heating from dust grains is also outlined in this section. Our model results are presented in Section~3, while a discussion of these and our final conclusions are given in Section~4. 

\section{Photoionization Modelling}

\subsection{Preliminary modelling}

We have constructed a photoionization model for knot J3 of Abell~30 based on the observational data of \citet{wesson03}, in order to investigate the temperature and ionic structures which could give rise to the observed ORLs and CELs. Given the very low surface brightness of the main extended nebula, and hence the very low density of material there, the effects of contamination of the spectra of the knots from this region are expected to be very small. Nevertheless, a background nebular spectrum was subtracted from the spectra of the knots by \citet{wesson03} in order to correct for any contamination. The three-dimensional photoionization code, {\sc mocassin}, described by \citet{ercolano03I}, was used in this work, since this code can self-consistently treat asymmetries and density and chemical inhomogeneities. Our model consisted of a three-dimensional Cartesian grid with the ionizing radiation source placed in one corner, with variable spatial sampling, with most of the cells being in the knot region, while the rest of the nebular region was assumed to be empty. 

The general procedure used to constrain the photoionization models descibed in this paper consists of an iterative process, involving the comparison of the predicted emission line fluxes, for some significant lines in the spectrum, with the values measured from the observations. The free parameters used in our models included nebular parameters, such as gas density and abundances, stellar parameters, such as luminosity, effective temperature and gravity, and, for the models described in Section~\ref{sub:dustheated}, dust parameters, such as dust-to-gas-ratio and grain size. The first step of the modelling process is to get a handle on the total gas mass and stellar luminosity required to match the total flux in the reference emission line as well as the relative intensities of the strongest lines in the spectrum. In hydrogen-rich environments the H$\beta$ line is generally chosen as the reference. However, in the case of knot J3 of Abell~30 we have chosen the He~{\sc i}~5876~{\AA} line as the reference, since helium is the most abundant element. The exact match will, of course, also depend upon the ionization structure of the nebula, which is dependent on the gas density and also on the central star temperature. In general, all model parameters are dependent on each other and it is not possible to completely isolate the effect of any given parameter on the resulting nebular structure and, consequently, on the nebular emission line spectrum. By looking at the diagnostic line ratios one can verify whether the temperature stucture of a given model is a realistic representation of the object being studied. The general ionization structure of the nebular gas is heavily dependent on the ionizing star parameters, as well as on the gas density distribution. The effective temperature and gravity of the central star were varied until the He$^{2+}$/He${^+}$ model ratio matched the value derived by the empirical analysis. The starting values were chosen to be L$_*$~=~5000\,$L_{\odot}$, T$_{eff}$~=~110,000\,K and log~$g$~=~6.0 \citep{harrington84}. Hydrogen-rich non-LTE model atmospheres were used for our preliminary modelling, however for the later and final models we used the non-LTE hydrogen deficient atmospheres described in Section~\ref{sub:modelatmospheres}.

In the first model constructed, the density and abundances in the knot were assumed to be homogeneous. A first attempt was made by using a density of 3200~cm$^{-3}$, as inferred from the [O~{\sc ii}] 3726~{\AA}/3729{\AA} line ratio \citep{wesson03}. Models were run both using the abundances derived from their empirical CEL analysis and those derived from the ORL analysis. Moreover the gas density was gradually scaled up until the observed He~{\sc i}~5876~{\AA} line flux was matched by the model. This approach, however did not succeed in reproducing the observed emission line spectrum, since the heating, which, in the virtual absence of hydrogen, mainly comes from helium photoelectrons, was not adequate. This resulted in the knot being too cold ($\sim$300~K) to collisionally excite the optical and ultraviolet heavy element forbidden lines observed in the spectrum. A similar problem was encountered by \citet{harrington84} when they tried to construct photoionization models of the hydrogen-deficient knots in Abell~30. This led them to hypothesize a chemical inhomogeneity in the knot, as mentioned in Section~\ref{sec:abell30intro}, with a central region where the ratio of C, N and O relative to He is very high ($\sim$10\%). However, as it was difficult to construct a one-dimensional photoionization model capable of self-consistently reproducing the postulated chemical inhomogeneity, Harrington \& Feibelman  could not test their hypothesis. 

\begin{table}
\begin{center}
\caption[Elemental abundance in knot J3 of Abell~30]{Model elemental abundances in the two assumed regions (dense, cold core and thin hotter outer region) of the knot J3 in Abell~30. Also listed, in columns~3 and~6, respectively, are empirical elemental abundances from the whole knot, derived from ORLs and from CELs \citet{wesson03}. The abundances are given by number with respect to helium.}
\label{tab:knotabun}
\begin{tabular}{lcccc}
\multicolumn{5}{c}{} \\
\hline
	& Core	 & ORLs   & Env. 		& CELs	\\
\hline
H/He	& 0.0250 & 0.0850 & 0.025  		& 0.085$^*$ \\
C/He	& 0.0200 & 0.0389 & 1.25$\cdot10^{-3}$ 	& 1.41$\cdot10^{-4}$  \\
N/He	& 0.0222 & 0.0229 & 1.87$\cdot10^{-3}$ 	& 6.76$\cdot10^{-5}$  \\
O/He	& 0.0233 & 0.1071 & 3.50$\cdot10^{-3}$ 	& 1.78$\cdot10^{-4}$  \\
Ne/He 	& 0.0200 & 0.0831 & 4.00$\cdot10^{-3}$ 	& 5.13$\cdot10^{-4}$  \\
\hline
\multicolumn{5}{c}{} \\
\end{tabular}
\\
\small{$^*$ No estimate for the helium abundance was available to \citet{wesson03} from CELs. H/He was assumed to be equal to the value they obtained from the ORL analysis. }
\end{center}
\end{table}

\begin{table}
\begin{center}
\caption{Adopted parameters for the final Abell~30 knot J3 photoionization model}
\label{tab:abelpar}
\begin{tabular}{lc|lc}
\multicolumn{4}{c}{} \\
\hline
log($L/L_{\odot}$)& 3.70  	  & R$_{core}$  	& 0.45$\cdot$R$_{knot}$ 	\\
T$_*$		& 130,000 K 		  & N$_{He}$(core) 	& 10,400 cm$^{-3}$ 		\\
log g		& 6.0			  & N$_{He}$(env)	& 1680 cm$^{-3}$  		\\
D$_{knot}$	& 1.68$\cdot$10$^{17}$cm & $\rho_d$/$\rho_g$(core/env) & 0.077/0.107 \\
R$_{knot}$ 	& 4.50$\cdot$10$^{15}$cm   & a$_d$		& 1.0$\cdot$10$^{-6}$cm  	\\
\hline
\end{tabular}
\end{center}
\small{	D$_{knot}$: distance of the centre of the knot from the ionizing source \\
	R$_{knot}$: radius of the hydrogen-deficient knot\\
	R$_{core}$: the radius of the thick, cold inner core of the knot\\
	N$_{He}$(core): helium density by number [cm$^{-3}$] in the inner core of the knot \\
	N$_{He}$(env): helium density by number [cm$^{-3}$] in the outer region of the knot \\
	$\rho_d$/$\rho_g$: dust-to-gas ratio by mass in the core and in the envelope of the knot\\
	a$_d$: radius of the dust grains}
\end{table}

With their result in mind, we constructed a second model, consisting of a spherical blob with two density/composition phases, where the core is assumed to be much denser and more metal-rich then the surrounding envelope. The abundances derived by \citet{wesson03} from the CELs were taken as an initial guess for the model abundances in the envelope of the knot and the abundances from the ORLs were taken as a starting point for the abundances in the core. However, the He/H abundance ratio cannot be derived from the observed CELs. This was set to a starting value of 0.5 by number, i.e. enhanced by a factor of 5 with respect to typical values for other PNe, the same factor by which the C/H abundance appeared to be enhanced from \citet{wesson03}'s CEL analysis. The abundances were iteratively adjusted as we tried to fit the observed emission line spectrum. Ultimately, this two-phase model also failed to produce enough heating for the envelope of the knot, which, according to Wesson et al.'s analysis, has an electron temperature of $T_{\rm e}$(OIII)~=~16680~K. 

\subsection{Dust heated models}
\label{sub:dustheated}

The failure of the first two models shows that photoionization of He and heavy elements alone cannot, in the virtual absence of hydrogen, provide enough heating to account for the collisionally excited emission lines observed, so an extra source of energy must be responsible for the heating necessary to produce the observed CELs. \citet{harrington84} proposed that the extra heating source might be provided by electron conduction from the shocked stellar wind. However, \citet{harrington96} later argued that, while this is a possibility, the presence of even a weak magnetic field could prevent such conduction and that a better candidate for the additional energy source, needed to produce the extra heating, is photoelectric emission from dust grains. This was first proposed by \citet{borkowski91} in their analysis of the hydrogen-deficient planetary nebula IRAS~18333-2357. Evidence for the presence of dust in Abell~30 is provided by the fact that \citet{cohen74} detected 10- and 20-$\mu$m emission from warm dust located within a radius of 5.5~arcsec of the central star, while \citet{cohen77} detected 1.6-3.5-$\mu$m emission from hot dust grains located in the same region. In addition, the central star appears to be reddened, and, as discussed by \citet{harrington84}, given the high galactic latitude of this object, the reddening is likely to be intrinsic. A study by \citet{borkowski94}, based on optical and near infrared images, confirmed the presence of dust in an expanding equatorial ring of H-poor gas around Abell 30; they detected no emission from the H-poor polar knots, concluding that there must be a dust deficiency there relative to the equatorial ring. This was confirmed by a more recent study by \citet{harrington97}, which showed the strength of the CELs relative to the He recombination lines to be much higher in the equatorial ring knots. They speculated that heating by photoionization might suffice in the polar knots, but that grain heating had to be dominant in the equatorial knots. The results of this paper show that dust heating is in fact needed in the polar knots, in order to reproduce the correct line intensities for CELs. The iplication of the dust deficiency of the polar knots to the equatorial knots and the results of the current work is that the dust-togas ratio in the equatorial knots must be much higher than that found here for the polar knots. In the recent analysis of \citet{wesson03}, on the other hand, it was found that the extinction coefficient, $c(H\beta)$, derived from the spectrum of the central star ($c(H\beta)$~=~0.60) and that derived for the knot J3 ($c(H\beta)$~=~0.55) are in good agreement. It is hard, however, to estimate the amount of dust actually locked into these  knots, and, therefore, in this work, we treated the dust-to-gas ratio as a free parameter. A radiative transfer treatment for dust grains has not yet been implemented in the radiative transfer of {\sc mocassin}, although work is already in progress in this direction. Nevertheless, in this work we estimated the heating provided by photoelectric emission from the grains by using a strategy similar to that already used by \citet{borkowski91} for the modelling of the grain-heated, dusty planetary nebula in M22. An outline of the method used in this work is given in the next section. 

The final model consisted of a spherical knot composed of two phases: a dense inner nucleus ($N_{\rm He}$(core)~$\approx$~10,500~cm$^{-3}$), with a radius equal to 0.45 of the total radius of the knot, $R_{\rm knot}$ (where  $R_{\rm knot}$\,=\,4.50$\cdot$10$^{15}$cm), surrounded by a less dense outer region ($N_{\rm He}$(env)~$\approx$~1700~cm$^{-3}$), occupying the remaining volume of the knot. Although the central core region only occupies about 9.1\% of the total volume of the knot, its higher density and higher metal abundance means that a significant fraction of the total mass of the knot (43.9\%) resides in this volume. As for the preliminary modelling the starting abundances were taken to be the empirical values obtained from CELs and from ORLs for the envelope and the core regions of the knot, respectively. The abundnaces were then adjusted, together with the other model parameters, in an iterative fashion in order to fit the the onbserved emission line spectrum. The final elemental abundances adopted for the two regions are summarised in Table~\ref{tab:knotabun}, together with the empirical abundances derived by \citet{wesson03}. The abundances for the outer region and for the cold dense core, where most of the ORL emission is produced, are given by number with respect to helium (columns~4 and 2, respectively, in Table~\ref{tab:knotabun}), since this is the most abundant element. The abundances in the core are compared to the empirical ORL abundances of \citet{wesson03}, also given by number with respect to helium in columns~2 and 3 of Table~\ref{tab:knotabun}. Most of the CEL emission is believed to be produced in the hotter envelope of the knot and, therefore, the model abundances in this region (column~4 of Table~\ref{tab:knotabun}) are compared with the empirical CEL abundances (column~5 of Table~\ref{tab:knotabun}). It is clear from Table~\ref{tab:knotabun}, however, that the final model abundances are quite different from the starting values which were taken to be equal to the values derived empirically. It should be noted at this point that, in Wesson et al.'s analysis, the abundances were derived without taking into account the possibility of chemical and density inhomogeneities within the knot, which would give rise to regions with different ionization structures. This implies a certain degree of freedom in the choice of abundances for our photoionization model. In the light of this, the discrepancies existing between model and empirical abundance values are easily explained. We discuss in Section~4 the implications of our model abundances in the light of Wesson et al.'s empirical analysis.

Table~\ref{tab:abelpar} lists the parameters used for the final simulation. We found that the ionization structure of this nebula was best reproduced by using a H-deficient model atmosphere for the ionizing source, with a stellar luminosity of L$_*$~=~4965~L$_{\odot}$, an effectice temperature of T$_*$~=~130,000~K and gravity defined by log~g~=~6.0. These parameters are in fair agreement with the values of T$_{eff}$~=~110,000~K, log~g~=~6.0 and L$_*$~=~5000~L$_{\odot}$ derived by \citet{harrington84} using the Zanstra method, and also with the results of Leuenhagen et al\@. (1993), who derived values of $T_{\rm eff}$~=~115,000~K and log~$g$~=~5.5 from their spectral analysis of the central star of Abell~30. The hydrogen-deficient non-LTE stellar atmospheres used in this work were calculated by using the T\"ubingen NLTE Model Atmosphere Package, as discussed in Section~\ref{sub:modelatmospheres}. Both the dust-to-gas ratio and the grain size have a direct influence on the amount of heating produced by the photoelectric emission process.  It is worth noting at this point that $\rho_d/\rho_g$ was treated as a free parameter, and different ratios were tried for the two regions of the knot, until the heating rate was high enough to reproduce the observed spectrum. 

\subsection{Grain Photoemission}
\label{sub:grains}
Photoelectric emission from dust grains occurs when electrons are ejected from a grain following the absorption of a photon. The ratio between the number of photoelectrons emitted and the number of absorbed photons defines the photoelectric yield, $Y$. This, of course, depends on the grain species and size, since a photoelectron could be excited by a photon several layers from the surface on the grain. In small grains the probability of the photoelectron being excited near the surface is higher than in larger grains, and, therefore, there is also a higher probability of the photoelectron escaping instead of being reabsorbed somewhere else inside the same grain. Experimental data exist for several materials \citep[e.g. the data for graphite and vitreous carbon by ][]{feuerbacher72}, but for the present calculations the following approximation for the photoelectric yield is used \citep{draine78}

\begin{eqnarray}
Y(h\nu)  = Y_{\infty}\left(1-\frac{B}{h\nu}\right)~~~~~~~~~~~~&~&~h\nu{\geq}B\\ \nonumber
Y(h\nu)  = 0~~~~~~~~~~~~~~~~~~~~~~~~~~~~~~~~~~~&~&~h\nu<B
\label{eq:y}
\end{eqnarray}

\noindent where B is a parameter which is chosen to be some energy above the photoelectric emission threshold energy, or work function, E$_w$, below which the yield becomes negligible, and Y$_{\infty}$ is another parameter which depends on the material used and is determined experimentally. In the {\it standard model} defined by \citet{draine78}, based on the results of \citet{feuerbacher72} and \citet{feuerbacheretal72}, which is appropriate for organic, metallic and silicate grains, Y$_{\infty}$~=~0.5 and B~=~8\,eV, where Draine's small grain size yield enhancement factors have been incorporated.  

Another important quantity which must be considered in estimating the grain heating is the energy distribution of photoelectrons ejected from the dust grains. In this calculation the photoelectron distribution function, $f(h{\nu},E)$, where $E$ is the photoelectron energy, is approximated using the formulation of \citet{borkowski91} for carbon grains, valid for photon energies above 11\,eV,

\begin{equation}
f(h\nu,E)\propto\begin{array}
{l@{\quad:\quad}l}
E & E<2.7\,{\rm eV} \\
{\left[E-1.7+\alpha^3(E-2.7)^3\right]^{-1}} & E>2.7\,{\rm eV}
\end{array}
\label{eq:f}
\end{equation}

\noindent where $\alpha$~=~0.059\,(26.4\,-\,$h\nu$) for $h\nu~<$~26.4\,eV and $\alpha$~=~0 for higher photon energies. For a particular photon energy, $h\nu$, this photoelectron energy distribution is truncated at an electron energy of $h\nu~-~E_w$, where E$_w$~=~5\,eV \citep{feuerbacher72}. For photon energies below 11\,eV, equation~\ref{eq:f} is scaled by replacing $E$ by $(11~-~E_w)\,E~/~(h\nu~-~E_w)$. This scaling is necessary because, from the experimental data, it is apparent that there is a shift in the energy at which the photoelectron energy distribution function, $f$, peaks, from $\sim$2.7\,eV at high incident photon energies, to lower electron energies for $h\nu~<$~11\,eV. In practice, $f(h{\nu},E)$ is the fraction of all electrons ejected by a photon of frequency $\nu$ which emerge with kinetic energies between $E$ and $E~+~dE$. 

In this work only a qualitative description of the role of dust in the energy balance of the nebula is adopted. We use a simplified approach, where grain charge and other factors are not taken into account, and the frequency-dependent contribution of the dust grains to the heating of the nebula, $G_d(\nu)$, is estimated at each grid cell according to

\begin{equation}
G_d(\nu)~=~n_d\,\cdot\,(\pi~a_d^2)\,\cdot\,F(\nu)\,\cdot\,Y(\nu)\,\cdot\,<E_{\nu}>
\label{eq:ggnu}
\end{equation}

\noindent where $n_d$ is the local dust density by number, $a_d$ is the radius of the dust grains assumed to be constant throughout the ionized region, and $F(\nu)$ is the photon flux. The quantity $<E_{\nu}>$ is the mean photoelectron energy, weighted by the photoelectron energy distribution function, $f(h\nu,E)$, namely, 

\begin{equation}
<E_{\nu}>~=~\frac{\int_{0}^{h\nu-E_w}f(h\nu,E)\,E\,dE}{\int_{0}^{h\nu-E_w}f(h\nu,E)\,dE}
\label{eq:enu}
\end{equation}

The total dust contribution to the heating is then obtained by integration of equation~\ref{eq:ggnu} over all photon energies above the work function, $E_w~=~h\nu_{th}$,  

\begin{equation}
G_d~=~\int_{\nu_{th}}^{\nu_{max}} G_d(\nu)\,d\nu
\label{eq:gd}
\end{equation}

\noindent where $\nu_{max}$ is the higher limit of the frequency range considered; in the {\sc mocassin} models of Abell~30, $\nu_{max}$ was set to 24~Ryd ($\approx$~326.4~eV). 

Finally, the total dust contribution to the heating, calculated at each grid cell, is added to the local thermal balance equations as an extra heating channel, so that the equation of thermal balance simply becomes

\begin{equation} 
G_{phot} + G_d = L_c +L_r+L_{ff}
\label{eq:thermaldust}
\end{equation}
\noindent where $G_{phot}$ is the heating due to the photoionization of the gas (mainly helium, in this case, but all other elements are also taken into account), $L_c$ is the contribution due to collisionally excited radiative cooling and $L_r$ and $L_{ff}$ are the contributions due recombination and free-free radiation, respectively. 

\subsection{Synthetic ionizing spectrum of the central star of \object{Abell~30}}
\label{sub:modelatmospheres}
For the central star it is more realistic to use NLTE
model atmosphere fluxes instead of blackbody ``spectra'' or simple
H+He composed models \citep{rauch02}. Consequently, we use synthetic spectra which have been
calculated from line-blanketed NLTE model atmospheres \citep{werner86, rauch97}.
These are plane-parallel and in hydrostatic and radiative equilibrium.

\begin{figure*}
\begin{center}
\epsfig{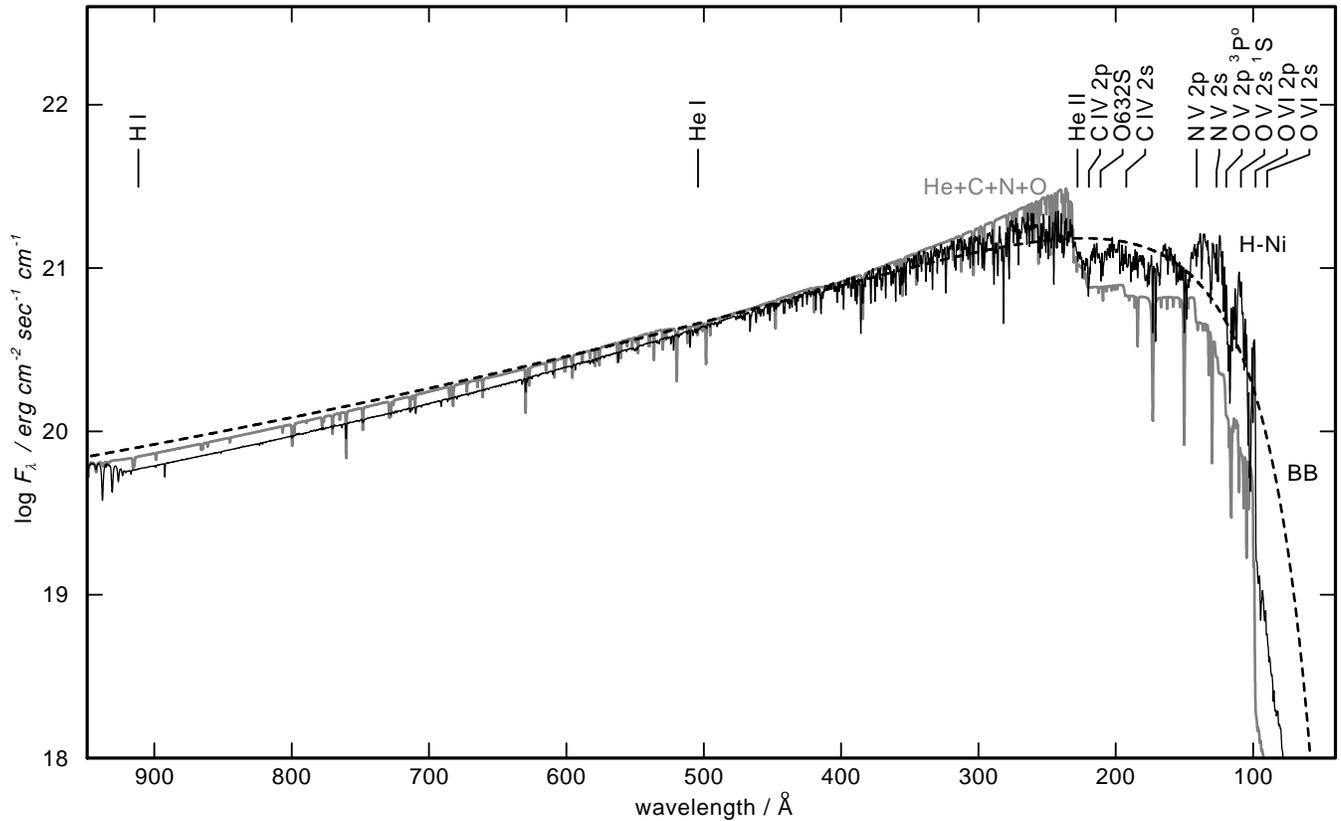}
\caption{Comparison of two synthetic NLTE model atmosphere fluxes calculated with
$T_{\mathrm eff} = 130,000\,\mathrm{K}$, $\log g = 6$, and different
chemical composition (Thin black line: H-rich model with normal cosmic abundances --labelled 'H-Ni'--; grey line: He+C+N+O model with an abundance ratio of He:C:N:O=41:40:4:15 by mass). 
Prominent absorption thresholds in the He+C+N+O spectrum are marked. The dashed line (labelled 'BB') denotes the flux of a black body with $T~=~130,000\mathrm{K}$.}
\label{fig:csfluxes}
\end{center}
\end{figure*}

\citet{leuenhagen93} presented a spectral analysis of the central star of Abell~30. They
found \Teffw{115}, \loggw{5.5}, He:C:N:O=41:40:4:15 by mass,
and $\log \dot{M}/\mathrm{M_\odot yr^{-1}} = -7.3$. Wind effects may have an influence
on the flux level in the UV \citep{Gabler89}. However, since the mass-loss
rate is relatively low, these will be neglected in our analysis.

We have calculated He+C+N+O models with \Teffw{110 - 150}, \loggw{6}
(cgs), and an abundance ratio of He:C:N:O=41:40:4:15 by mass.
The statistics of the model atoms are summarized in Table\,\ref{stat}.  The observed nebular emission-line spectrum is best reproduced by our {\sc Mocassin}
    photoionization models when a synthetic spectrum
    calculated from our stellar He+C+N+O model atmosphere grid, with
    $T_\mathrm{eff} = 130,000\,\mathrm{K}$ and  $\log g = 6$ is employed
    as the ionizing source.
In Figure~\ref{fig:csfluxes} the fluxes obtained using a hydrogen-deficient model are compared to those obtained using a black body and a hydrogen-rich model with $T_{\mathrm eff} = 130,000\,\mathrm{K}$ and $\log g = 6$. A comparison between the nebular results obtained using a hydrogen-deficient and a hydrogen-rich model with normal cosmic abundances, for the same central star and nebular parameters is presented in Section~\ref{sub:differences}.

\begin{table}
\caption{Statistics of model atoms used in the calculation  
         of the NLTE stellar models. NLTE is the number of levels
         treated in NLTE, RBB is the number of line transitions.
         194 additional levels are treated in LTE.} \vspace{3mm}
\label{stat}
\begin{center}
\begin{tabular}{lcc}
\hline
ion & NLTE & RBB \\
\hline
\ion{He}{i}    &   1 &   0  \\
\ion{He}{ii}   &  14 &  78  \\
\ion{He}{iii}  &   1 &   -  \\
\ion{C }{iii}  &   3 &   1  \\
\ion{C }{iv}   &  54 & 288  \\
\ion{C }{v}    &   1 &   0  \\
\ion{N }{iv}   &   3 &   1  \\
\ion{N }{v}    &  44 & 150  \\
\ion{N }{vi}   &   1 &   0  \\
\ion{O }{v}    &   6 &   3  \\
\ion{O }{vi}   &  43 & 166  \\
\ion{O }{vii}  &   1 &   0  \\
\hline
\end{tabular}
\end{center}
\end{table}

\subsection{Modifications to the {\sc mocassin} code}
Some modifications to the original version of the {\sc mocassin} code had to be made in order to include the dust photoelectric heating treatment described in Section~\ref{sub:grains}. They mainly consisted of the addition of a dust heating term (Equation~\ref{eq:gd}) to the thermal balance equation (Equation~\ref{eq:thermaldust}). These modifications, however, will not be included in the public release version of {\sc mocassin}, which will consist of the pure photoionization version of the code only, since the intention is to enhance the code in the near future by including dust grains in the radiative transfer treatment. The analysis carried out in the current paper can, therefore, be considered to be a qualitative investigation of the effects of dust photoelectric heating in hydrogen-deficient environments.

\section{Abell~30 model results}

\subsection{The temperature and the ionization structure}

\renewcommand{\baselinestretch}{1.2}
\begin{table}
\begin{center}
\caption{Mean temperatures (K) weighted by ionic species for knot J3 of Abell~30. For each element the upper row is for the optically thick inner core and\ the lower row is for the optically thin envelope of the knot.}
\begin{tabular}{lcccccc}
\multicolumn{7}{c}{} \\
\hline
	&	&	&	& Ion	 &	&	\\
\cline{2-7}
El.	& {\sc i}	&{\sc ii}	&{\sc iii}&{\sc iv}&{\sc v}	&{\sc vi}\\
\hline
H	& 890		& 970	& 	&	&	&	\\
	& 17,270 	& 17,320 	& 	&  	& 	& 	\\
	&	&	&	&	&	&	\\
He	& 500		& 950		& 1,430	&  	& 	& 	\\
	& 10,540	& 15,550	& 17,660	&  	& 	& 	\\
	&	&	&	&	&	&	\\
C	& 630		& 750		& 970		& 1,200		& 1,520		& \\
	& 11,600	& 11,820	& 15,830	& 17,500	& 17,780	& \\
	&	&	&	&	&	&	\\
N	& 550		& 630		& 960		& 1,200		& 1,510		& 1,520	\\
	& 11,340	& 11,760	& 16,320	& 17,580	& 17,780	& 17,690\\
	&	&	&	&	&	&	\\
O	& 550		& 670		& 1,010		& 1,500		& 1,550		& 1,650 \\
	& 11,310	& 11,760	& 16,460	& 17,690	& 17,800	& 17,690\\
	&	&	&	&	&	&	\\
Ne	& 500		& 690		& 1,040		& 1,510 	& 1,550		& \\
	& 11,510	& 11,910	& 16,840	& 17,730	& 17,670	& \\
\hline
\label{tab:abeltemps}
\end{tabular}
\end{center}
\end{table}
\renewcommand{\baselinestretch}{1.5}

\renewcommand{\baselinestretch}{1.2}
\begin{table}
\begin{center}
\caption{Fractional ionic abundances for knot J3 of Abell~30. For each element the upper row is for the optically thick inner core and\ the lower row is for the optically thin envelope of the knot. Exponent powers of ten are given by the numbers in brackets. }
\begin{tabular}{lcccccc}
\multicolumn{7}{c}{}\\
\hline
	&	&	&	& Ion	 &	&	\\
\cline{2-7}
El.	& {\sc i}	&{\sc ii}	&{\sc iii}&{\sc iv}&{\sc v}	&{\sc vi}\\
\hline
H	& .19(-1)& .981	& 	& 	& 	& 	\\
	& .77(-3)& .999	& 	&  	& 	& 	\\
	&	&	&	&	&	&	\\
He	& .56(-1)& .904	& .39(-1)	&  	& 	& 	\\
	& .23(-3)& .161	& .837	&  	& 	& 	\\
	&	&	&	&	&	&	\\
C	& .66(-3)& .112	& .800	& .87(-1)& .24(-3)	&	\\
	& .27(-5)& .71(-2)	& .139	& .530	& .324	&	\\
	&	&	&	&	&	&	\\
N	& .67(-3)& .140& .620	& .238	& .61(-3)	& .16(-6)	\\
	& .27(-5)& .68(-2)& .191	& .694	& .106	& .18(-2)	\\
	&	&	&	&	&	&	\\
O	& .82(-3)& .158 	& .820	& .21(-1)	& .42(-4)	& .19(-8)	\\
	& .16(-5)& .66(-2)	& .278	& .627		& .89(-1)	& .17(-3)	\\
	&	&	&	&	&	&	\\
Ne	& .30(-2)& .183	& .811	& .28(-2)	& .50(-5) 	& 	\\
	& .78(-5)& .15(-2)	& .360	& .583		& .43(-1)	& 	\\
\hline
\label{tab:abelionratio}
\end{tabular}
\end{center}
\end{table}
\renewcommand{\baselinestretch}{1.5}

Table~\ref{tab:abeltemps} lists mean temperatures (given to an accuracy of four significant figures), weighted by the ionic abundances, for the two regions of the knot. The upper entries in each row of Table~\ref{tab:abeltemps} are for the inner core and the lower entries are  for the surrounding envelope. The value of T$_e$(O~{\sc iii})~=~16,460~K obtained by the {\sc mocassin} model for the outer region of the knot is in agreement with the empirical value derived from CELs by \citet{wesson03} of T$_e$(O~{\sc iii})~=~16,680~K, as is also confirmed by the [O~{\sc iii}]~(4959~{\AA}~+~5007~{\AA})/4363~{\AA} model and empirical ratios that are in good agreement. The temperature corresponding to the predicted [O~{\sc iii}]~(4959~{\AA}~+~5007~{\AA})/4363~{\AA} ratio of 52.4 is 17,000~K, which is about 500~K higher than the mean temperature weighted by the O~{\sc iii} abundance, as reported in Table~\ref{tab:abeltemps}.  This can be explained by the fact that there could be fluctuations in temperature in the O$^{2+}$ region, in which case the empirical temperature derived by the [O~{\sc iii}] diagnostic ratios would be weighted towards the higher limit of the range.

The volume-averaged  fractional ionic abundances for the two regions of the knot are listed in Table~\ref{tab:abelionratio}, where, once again, the upper entries for each element are for the inner core and the lower entries are for the surrounding envelope of the knot.

It can be seen from Tables~\ref{tab:abeltemps} and~\ref{tab:abelionratio} that the model predicts very different temperature and ionization structures for the two regions of the knot. This has important implications for empirical determinations of abundances from spectroscopic observations, which generally assume constant ionization fractions throughout a region. In this light it is possible to understand the differences between the elemental abundances inferred using {\sc mocassin} models and those obtained by \citet{wesson03} using empirical methods. 

\begin{figure}
\begin{center}
\epsfig{file=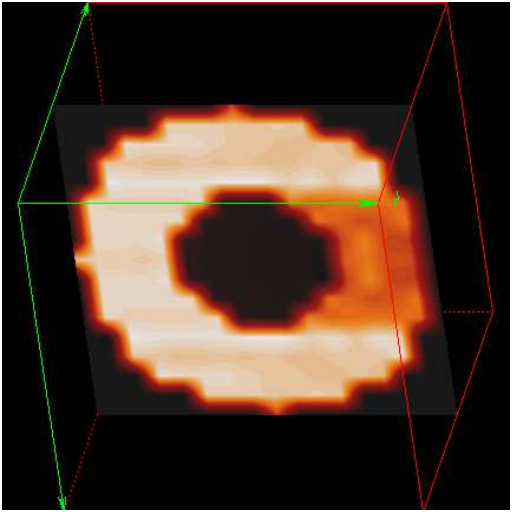,width=0.8\linewidth,clip=,bbllx=90, bburx=520,bblly=182,bbury=613,angle=-90.}
\end{center}
\caption[Predicted electron temperature distribution in knot J3 of Abell~30] {Cross-section map of the electron temperature distribution in the {\sc mocassin} model of knot J3 of Abell~30. The slice plot is obtained on a plane through the centre of the knot and it is oriented such that the ionizing stellar field is incoming from the left-hand side. }
\label{fig:abeltemps}
\end{figure}

Finally, Figure~\ref{fig:abeltemps} shows a {\it slice} plot of the temperature distribution in the knot. In this figure the knot is oriented such that the stellar ionizing radiation is coming from the left-hand side. The large temperature variations existing in this model are illustrated by this plot, as well as the cooler shadow region created by the thick inner core.

\subsection{The nebular emission line spectrum}

\begin{table}
\begin{center}
\caption{Dereddened observations and predicted emission lines fluxes for knot J3 of Abell~30. The line fluxes are given relative to He~{\sc i}~5876\,{\AA}, on a scale where He~{\sc i}~5876\,{\AA}~=~1. The flux of the He~{\sc i}~5876\,{\AA} line is also given in absolute units [10$^{-15}$erg\,cm$^{-2}$\,s$^{-1}$].}
\begin{tabular}{lcccc}
\multicolumn{5}{c}{}\\ 
\hline
Line			& \multicolumn{3}{c}{Predicted} & Observed$^a$ \\
\cline{2-5}
				& Core	& Envel	& Total & 		\\
\hline
He~{\sc i} 5876/10$^{-15}$$\frac{erg}{cm^{2}s}$ 
				& 7.07	& 0.158	& 7.23  & 7.37	\\
He~{\sc i} 5876			& 0.979	& 0.021	& 1. 	& 1.	\\
He~{\sc ii} 4686		& 1.24  & 0.869 & 2.11	& 2.50	\\ 
H$\beta$ 4861			& 0.062	& .0023	& 0.064	& 0.080 \\
C~{\sc ii} 4267			& 0.347	& .0002	& 0.347	& 0.452 \\
C~{\sc iii}$]$ 1908 		& 0.	& 1.52	& 1.52	& 2.8	\\
C~{\sc iii} 4648 		& 0.044	& 0.002	& 0.046	& 0.030	\\
C~{\sc iii} 2297		& 1.53	& 0.016	& 1.53	& 0.960	\\
C~{\sc iv} 1550$^b$		& 0.	& 15.42	& 15.42	& 9.2	\\
$[$N~{\sc ii}$]$ 6584		& 0.	& 0.141	& 0.141	& 0.829 \\
N~{\sc iv}$]$ 1486		& 0.	& 4.16	& 2.62	& 2.56	\\
$[$O~{\sc ii}$]$ 3727		& 0.030	& 0.120	& 0.150	& 1.66	\\
O~{\sc ii} 4303			& 0.030	& 0.	& 0.030	& 0.094 \\
O~{\sc ii} 4097$^c$		& 0.037	& 0.	& 0.037	& 0.174	\\
O~{\sc ii} 4089			& 0.072	& 0.	& 0.072	& 0.060	\\
O~{\sc ii} 4075			& 0.158	& 0.	& 0.158	& 0.149 \\
$[$O~{\sc iii}$]$ 5007		& 0.	& 10.7	& 10.7	& 14.2	\\
$[$O~{\sc iii}$]$ 4363		& 0.	& 0.275	& 0.275	& 0.275	\\
$[$Ne~{\sc iii}$]$ 3869		& 0.	& 7.57	& 7.57	& 2.85	\\
$[$Ne~{\sc iv}$]$ 2423		& 0. 	& 18.22	& 18.22	& 15.2	\\
$[$Ne~{\sc v}$]$ 3426		& 0. 	& 1.18	& 1.18	& 1.20	\\
\hline
\label{tab:abellines}
\end{tabular}
\end{center}
\small{$^a$~\citet{wesson03}.\\
       $^b$ Can be attenuated by dust absorption.\\
       $^c$ Observed line probably blended with the secondary Bowen fluorescence line from N~{\sc iii}.
          }
\end{table}

The adopted parameters for the models are those listed in Tables 1 and 2. The fluxes of some important emission lines predicted by {\sc mocassin} for the final model of knot J3 in Abell~30 are listed in Table~\ref{tab:abellines}, together with the corresponding observed values \citep{wesson03}. The fluxes are given relative to He~{\sc i}~5876~{\AA}, since helium is the most abundant element in this object. The contributions from each of the two assumed regions of the knot are given separately in the table and, as expected, it is clear that in our model the ORLs are mainly emitted in the cold inner core of the knot, while the CELs are emitted in the hotter envelope surrounding it. The predicted line fluxes are in satisfactory agreement with the observed values, although some discrepancies still remain, as discussed in the following paragraph. A dust to gas ratio in the envelope of 0.107 by mass was sufficient to produce a heating rate adequate for the production of the observed CELs. In the central core the dust-to-gas ratio was set to 0.077 by mass, although, as will be discussed later, photoelectric emission from dust grains is not the dominant heating mechanism in this region.  

The largest discrepancies are observed for the [O~{\sc ii}]~3727~{\AA} and [N~{\sc ii}]~6584~{\AA} lines. Both of these lines, however, may have a recombination component \citep{rubin86}. This component is accounted for by {\sc mocassin} in the case of [O~{\sc ii}]\,3727\,{\AA}, but not in the case of [N~{\sc ii}]\,6584\,{\AA}. The statistical equilibrium calculation routines were modified in order to self-consistently calculate the contribution of recombination to the population of each of the 5 levels considered in the O$^{2+}$ ion. As the changes to the statistical equilibrium routines were made globally, the contribution of recombination to any of the collisionally excited lines considered in a simulation can be taken into account, given that the required recombination coefficients are known and have been added to the atomic data files. In our model we used the recombination coefficients given by \citet{liu00}, their calculations however only covered the range of electron temperatures from 5,000~K to 20,000~K, while the electron temperature in the core of the knot (where most of the recombination lines are expected to be emitted) are in the range from 500~K to 2000~K. Low temperatures calculations for these recombination coefficients are not yet available, although work is being carried out at UCL by P. J. Storey. In the current work we have fitted a power law to the available data, in order to extrapolate the coefficients to the low temperatures required. The inclusion of radiative recombination contributions to the final flux of [O~{\sc ii}]\,3727\,{\AA} resulted in an increase of 25\% (from a value of 0.120 to a value of 0.150 with respect to He~{\sc i}\,5876\,{\AA}); this increase, although significant, is not sufficient to remedy the discrepancy with the observed value of 1.66 with respect to He~{\sc i}\,5876\,{\AA}. Due to the high electron densities in the core region of the knot, collisional de-excitation plays an important part in depopulating the 2p$^3$~$^2$D level of O$^{2+}$ hence suppressing the $^2$D$_{\frac{5}{2}, \frac{3}{2}}$-$^4$S$_{\frac{3}{2}}$ doublet transition at 3726,29~{\AA}. The same is believed to occur in the case of the [N~{\sc ii}]\,6584,48\,{\AA} transitions which are also underestimated in our models. An intermediate {\it transition} phase between the core and the envelope might provide the solution to this discrepancy; in our models we used a {\it clear-cut}, bi-density, bi-chemistry model, while a gradual decay in density and metal abundances might have been a more realistic description -- the lower densities in such a region would lead to stronger CEL emission. 

There are also discrepancies between {\sc mocassin}'s prediction of the fluxes for some of the optical recombination lines of O~{\sc ii} and the observations. However, O~{\sc ii}~4303~{\AA}, 4097~{\AA} and 4089~{\AA} all come from the 3d\,$^4$F level and therefore their relative strengths are only determined by the branching ratios, which are well known \citep[e.g., see ][]{liu95}. The observed relative intensities of these lines do not conform to the theoretical branching ratios. As indicated in Table~\ref{tab:abellines}, the O~{\sc ii}~4097~{\AA} line should not be taken into account, since this may be blended with one of the secondary Bowen fluorescence lines of N~{\sc iii}. Nevertheless, the relative fluxes of O~{\sc ii}~4303~{\AA} and O~{\sc ii}~4089~{\AA} are also in disagreement with the theoretical branching ratios. \citet{tsamis03} also noticed a similar effect in the spectra of five H~{\sc ii} regions, namely, M~17, NGC~3576, 30~Doradus, LMC~N11B and SMC~N66. They suggested that at the low electron densities characteristic of those nebulae, a breakdown of the statistical equilibrium may occur among the fine-structure levels of the parent $^3$P$_{0,1,2}$, ground term of recombining O$^{2+}$, causing the observed behaviour. However, in the case of the high electron density ($N_{\rm e}\sim$12,500~cm$^{-3}$) in the core region of the knot J3 of Abell~30, this is not expected to occur. 

\begin{table}
\begin{center}
\caption[Diagnostic ratios for electron temperature, $T_e$, and electron density, $N_e$ (Biconical model).] {Diagnostic ratios for electron temperature, $T_e$, and electron density, $N_e$. Ratios predicted by {\sc mocassin} (column~3) and observed ratios (column~5).}
\begin{tabular}{lccccc}
\multicolumn{4}{c}{} \\
\hline
Ion		& Lines 	& Ratio	 	& Ratio 	\\
		& ({\AA})	& ({\sc mocassin})	& (Observed)	\\	 
\hline								
$N_e$		&		& 		&		\\
\hline								
[O~{\sc ii}]	& 3726/3729	& 1.68		& 1.73		\\
\hline								
$T_e$		&		& 		&		\\
\hline								
He~{\sc i}	& 5876/4471	& 2.80		& 2.76		\\
He~{\sc i}      & 6678/4471	& 0.802		& 0.777		\\
O~{\sc ii}	& 4075/4089	& 2.19		& 2.47		\\
$[$O~{\sc iii}$]$   & (5007+4959)/4363 & 52.5	& 54.5	 	\\
\hline
\label{tab:diagnosticsabell}
\end{tabular}
\end{center}
\end{table}

Table~\ref{tab:diagnosticsabell} lists the diagnostic ratios for electron density and temperature predicted by {\sc mocassin} and those observed by \citet{wesson03}. Good agreement is obtained for most line ratios. A lower ratio (-11.3\%) was obtained for O~{\sc ii}~4075~{\AA}/4089~{\AA}, indicating that the {\sc mocassin} model predicts a slightly lower temperature (1,150~K) for the inner core of the knot than the value implied by the observed ratio (2,800~K). This could be adjusted by using a slightly higher value for the dust-to-gas ratio in the core region of the knot.

\subsection{The thermal balance}

\begin{figure}
\begin{center}
\psfig{file=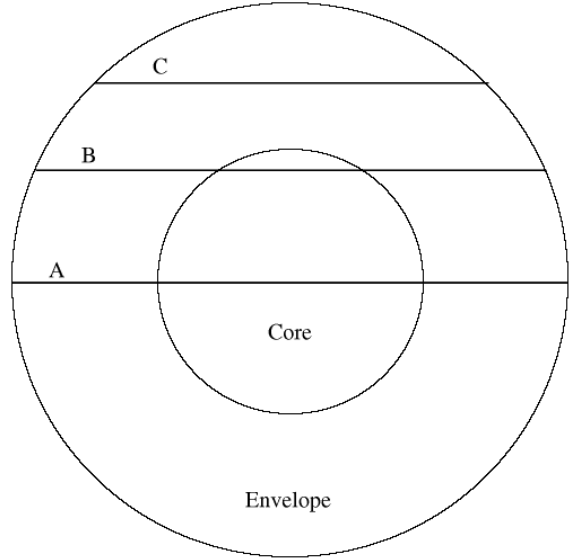, height=75mm, width=75mm}
\caption{Schematical representation of a cross-section through the centre of the knot. The three rays, labelled {\it A}, {\it B} and {\it C}, correspond to the positions for which the model thermal contributions are plotted in Figure~\ref{fig:heat}. The stellar field is assumed to be incoming from the left-hand side.}
\label{fig:circles}
\end{center}
\end{figure}

\begin{figure*}
\begin{center}
\begin{minipage}[t]{75mm} 
\psfig{file=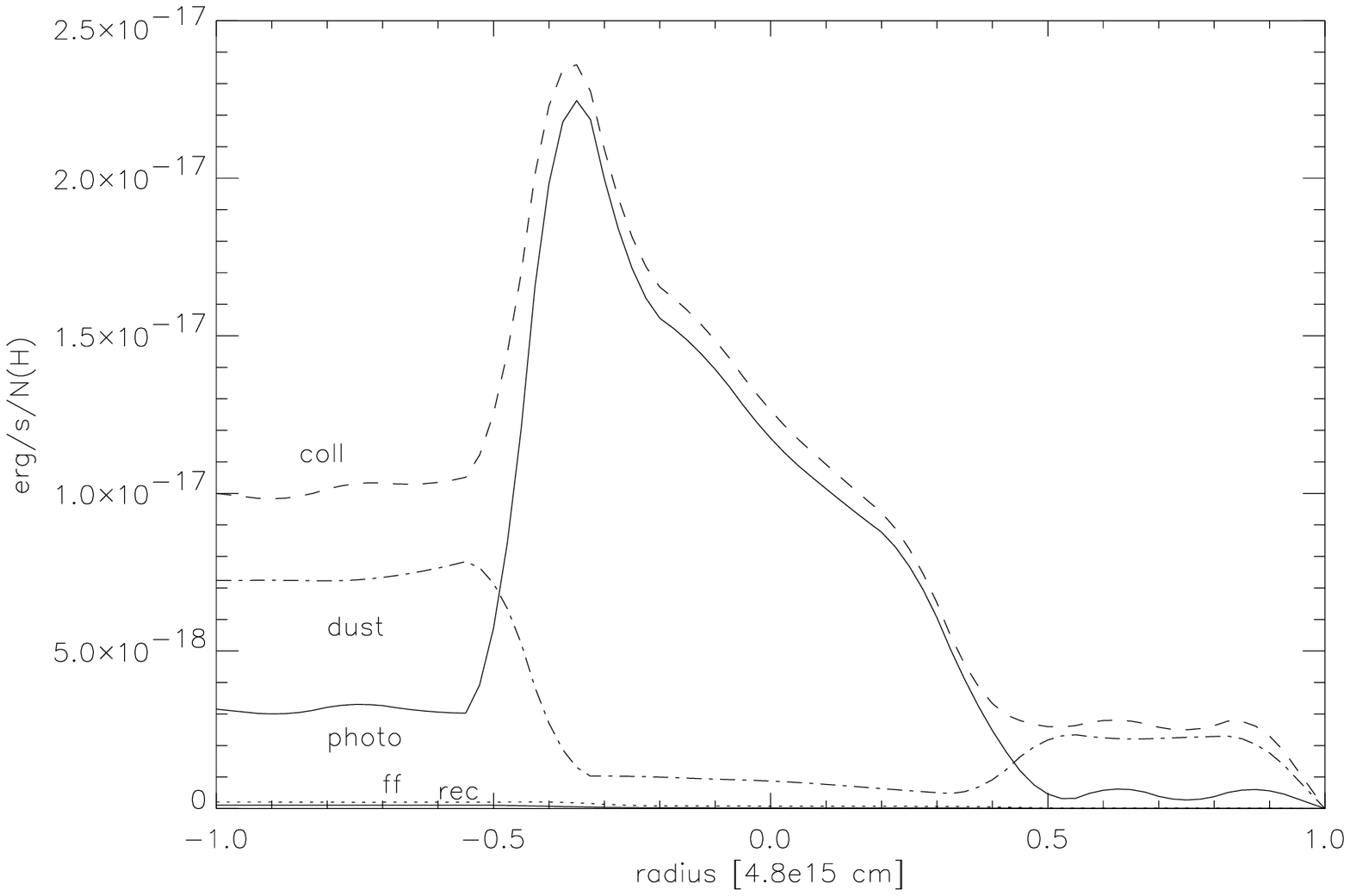, height=75mm, width=75mm}
\end{minipage}
\begin{minipage}[t]{75mm} 
\psfig{file=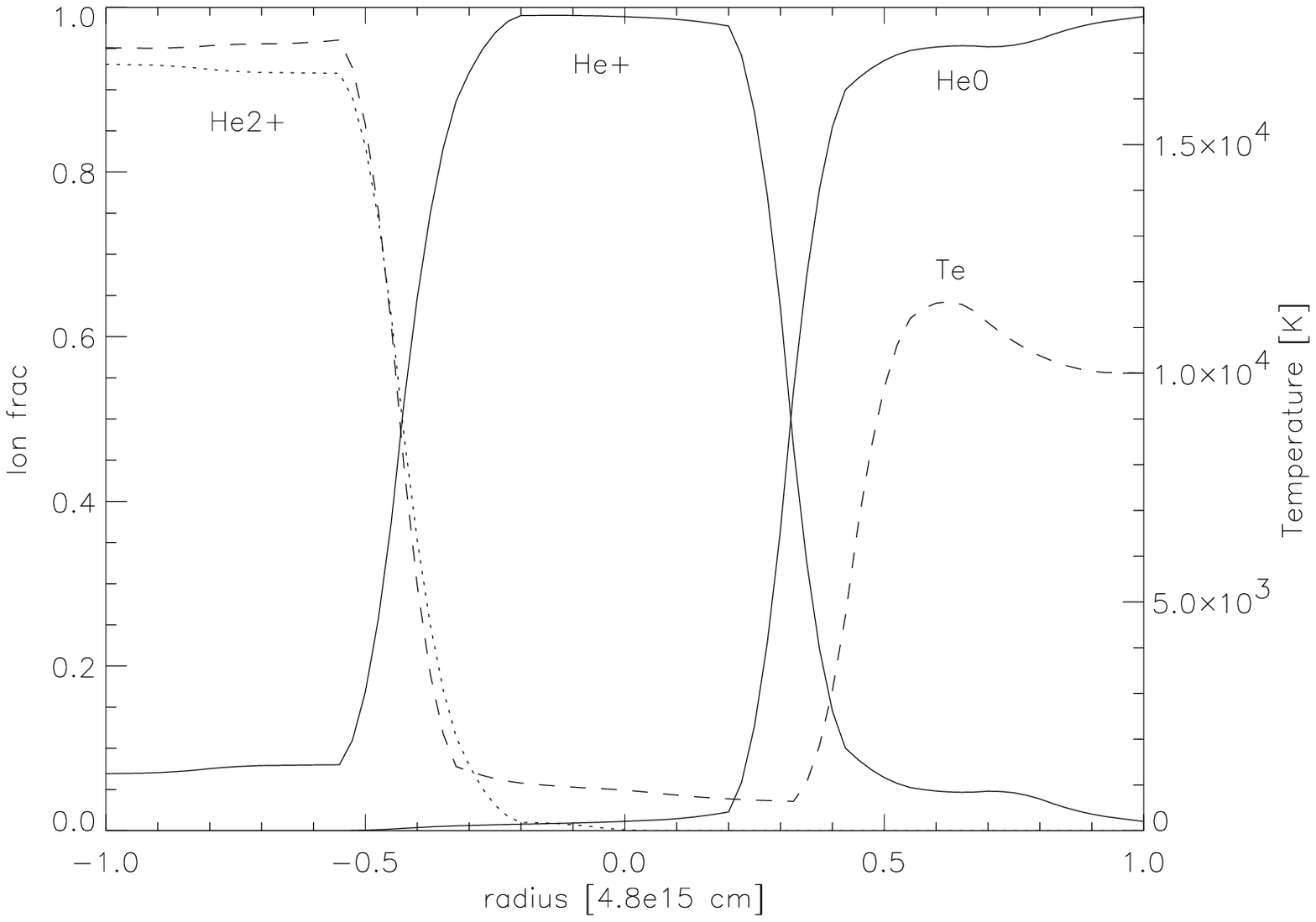, height=75mm, width=75mm}
\end{minipage}
\begin{minipage}[t]{75mm}
\psfig{file=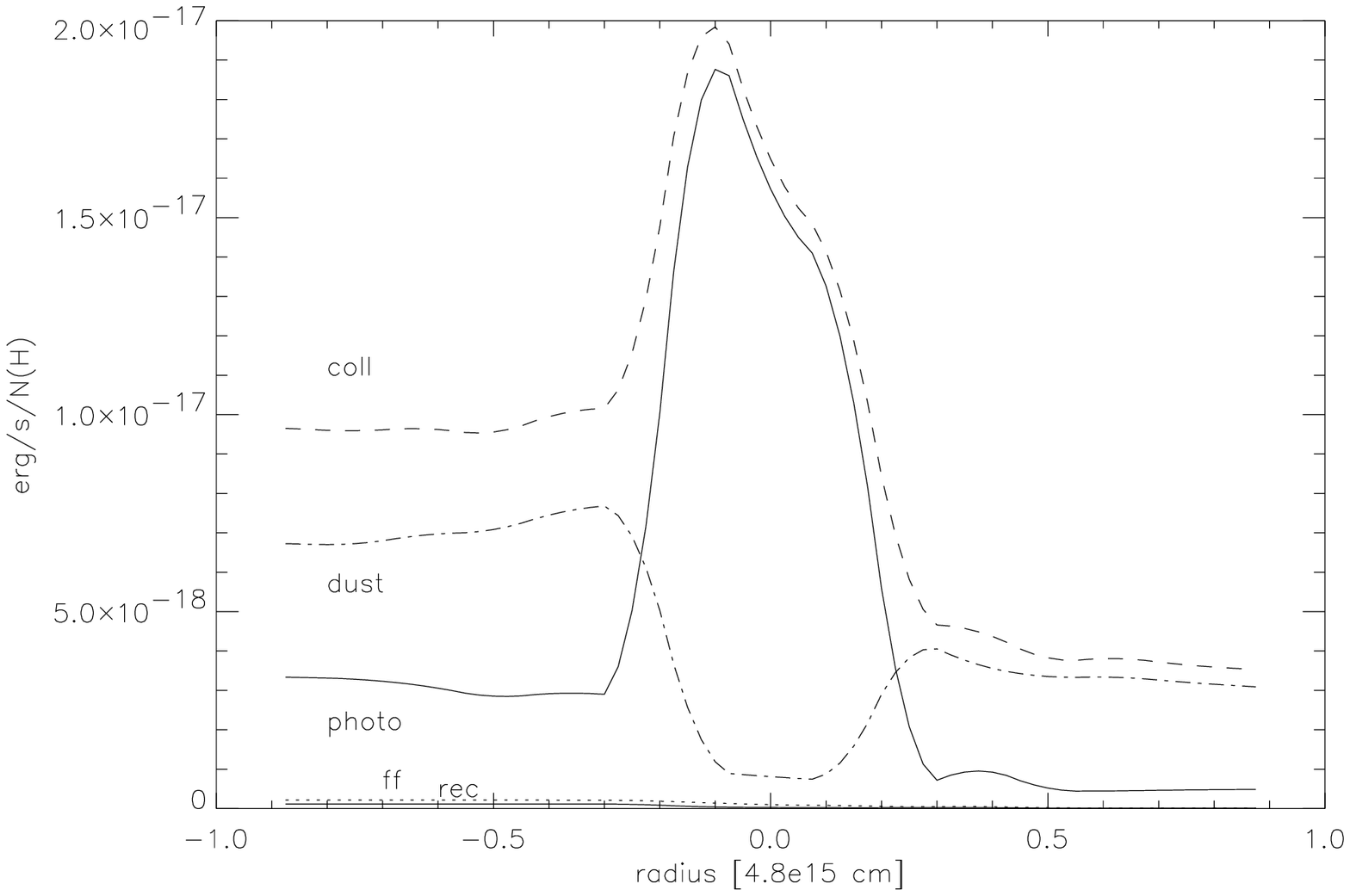, height=75mm, width=75mm}
\end{minipage}
\begin{minipage}[t]{75mm} 
\psfig{file=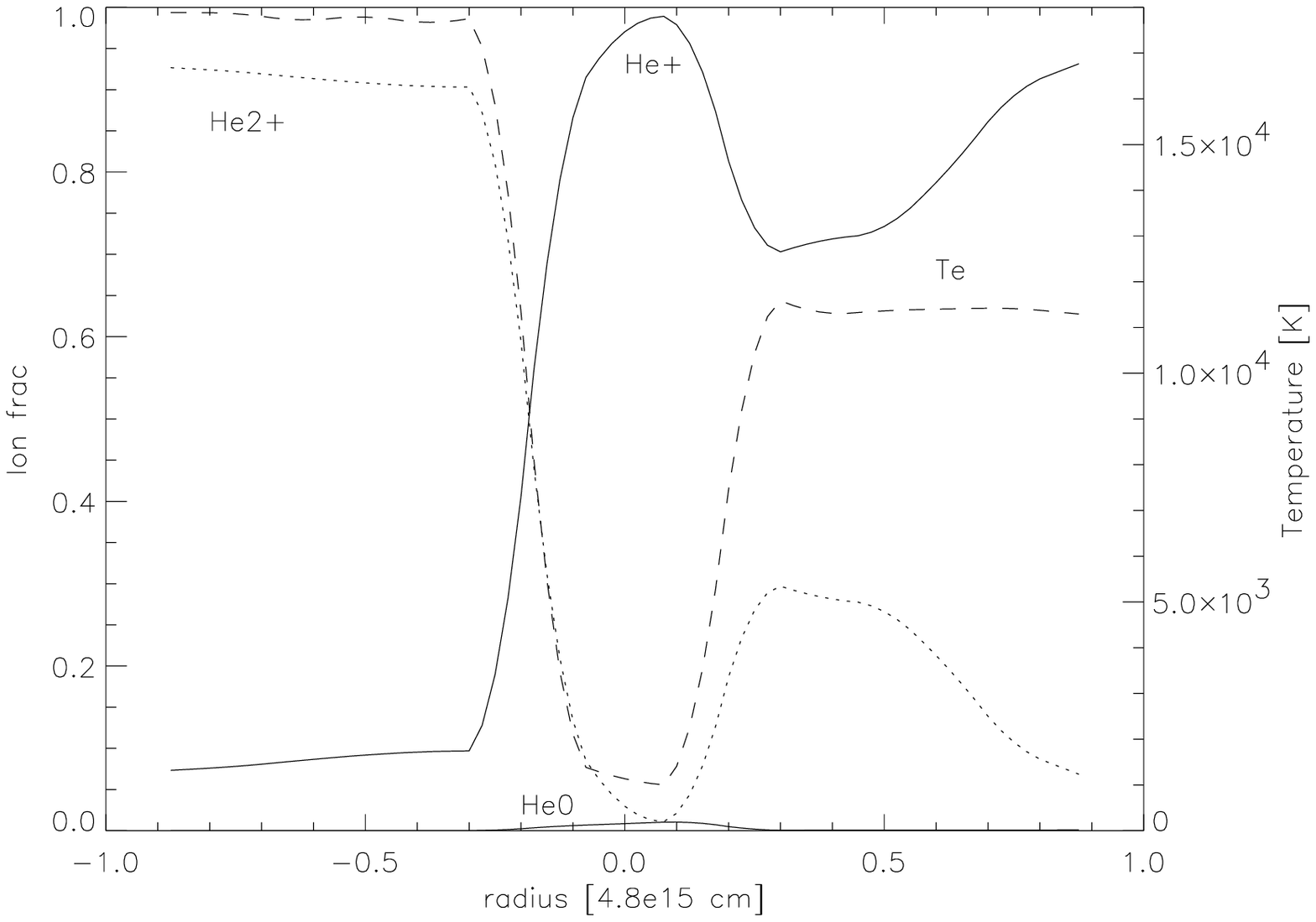, height=75mm, width=75mm}
\end{minipage}
\begin{minipage}[t]{75mm} 
\psfig{file=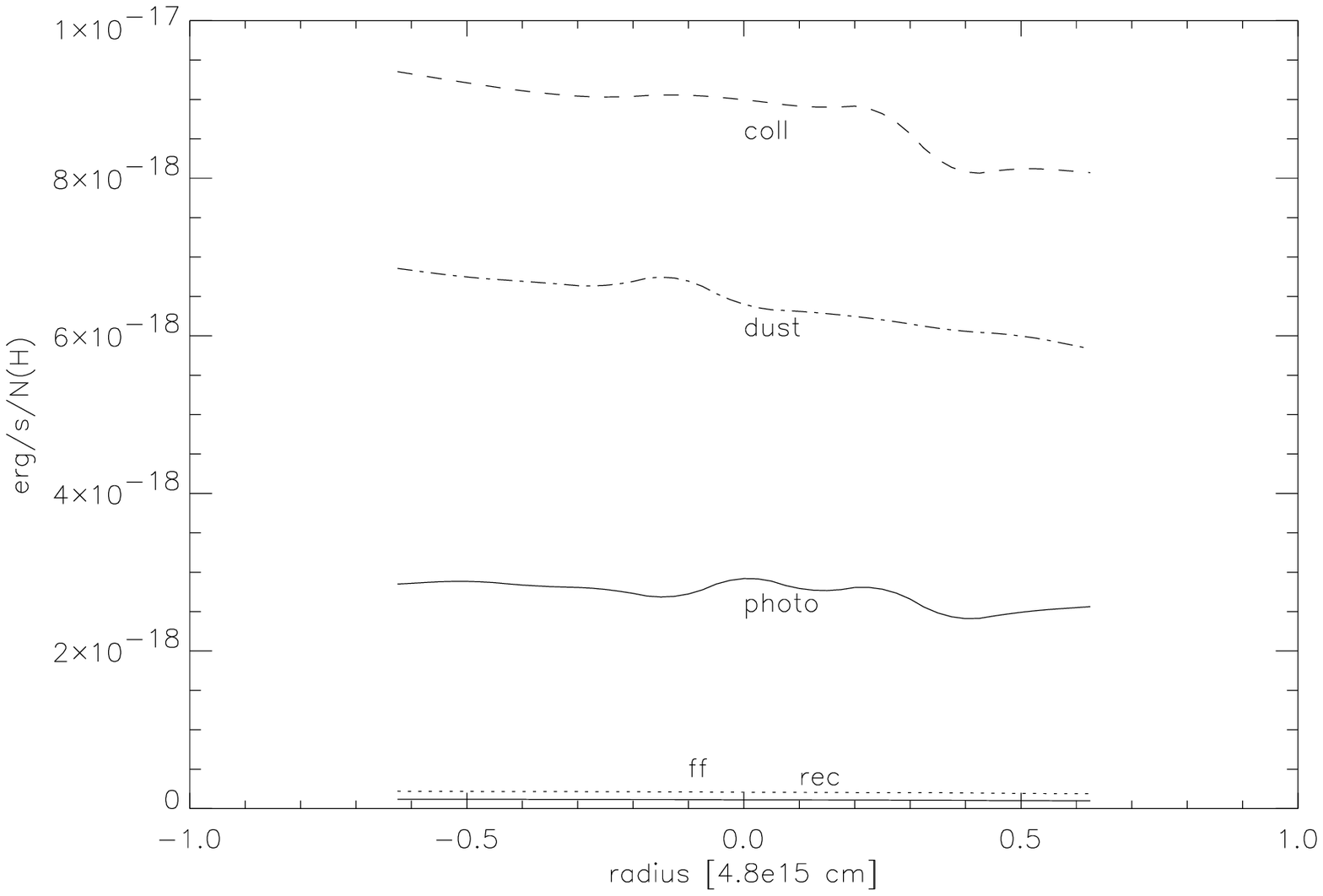, height=75mm, width=75mm}
\end{minipage}
\begin{minipage}[t]{75mm} 
\psfig{file=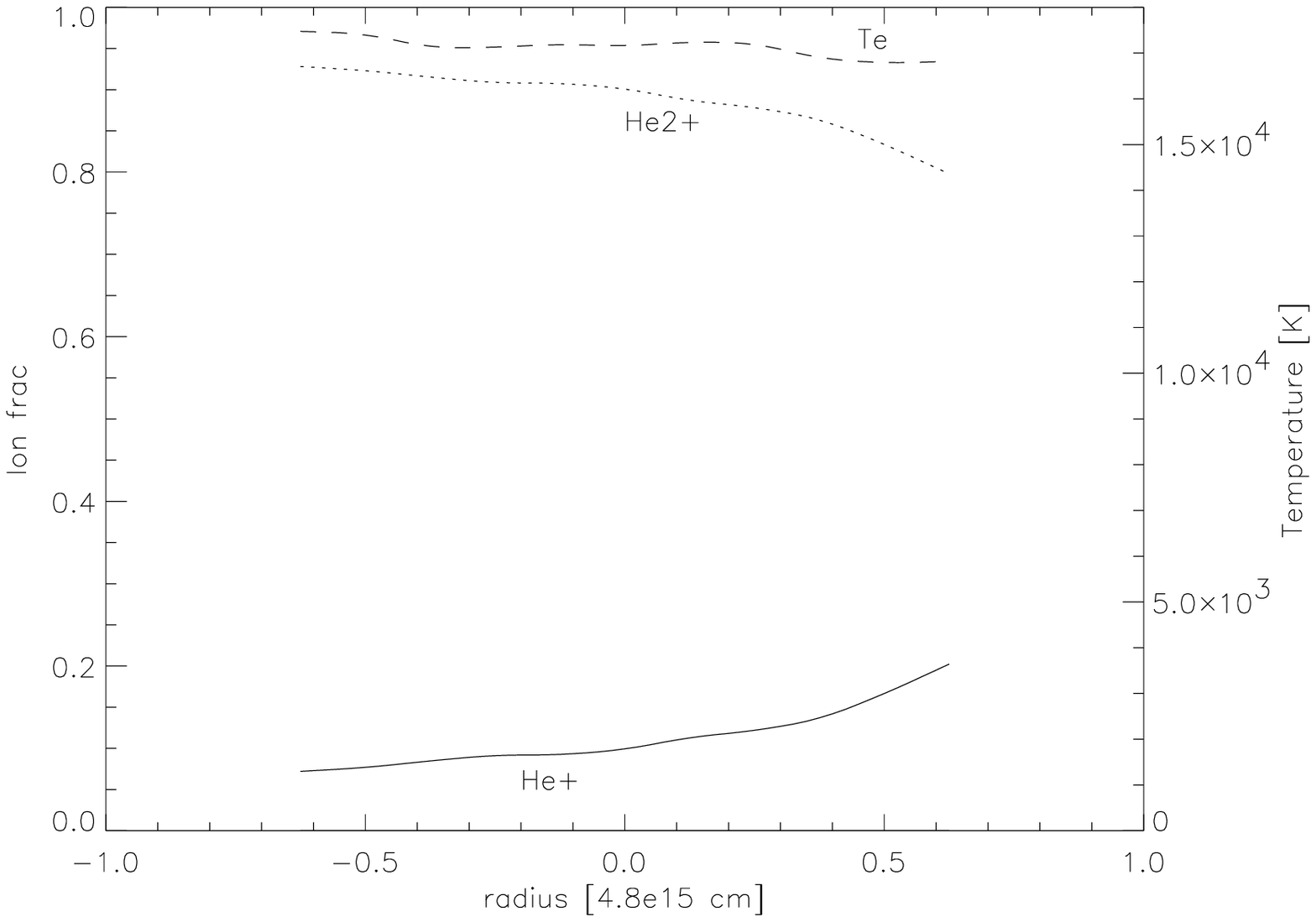, height=75mm, width=75mm}
\end{minipage}
\caption{{\it Left panels:} Heating and cooling contributions for knot J3 of Abell, plotted as a function of radius for three different cuts (from top to bottom: cuts A, B and C as shown in Figure~\ref{fig:circles}).({\it photo}: photoelectric heating by atoms and ions; {\it dust} photoelectric heating by dust grains; {\it coll}: collisional cooling by ions; {\it ff}: cooling by free-free radiation; {\it rec}: cooling by recombination).  {\it Right panels:} Corresponding fractional ionic abundances for helium and electron temperatures, plotted as a function of radius.}
\label{fig:heat}
\end{center}
\end{figure*}

From the analysis of the heating and cooling contributions obtained at each location in our model grid, it appears that dust grains play the leading role in the heating of the knot in the envelope, dominating over the heating produced by photoionization of helium, which is the most abundant element, and by other elements (including the small quantity of hydrogen remaining in the knot). This is very clearly visible in the left-hand panels of Figure~\ref{fig:heat}, where the heating and cooling contributions due to all the energy channels considered in our simulation are plotted as a function of radius. These include cooling by collisionally excited emission lines, cooling by free-free and recombination continuum radiation and heating by photoionization and photoelectric emission from dust grains. The corresponding helium ionic fractions and the electron temperatures are plotted on the right-hand panels of Figure~\ref{fig:heat}, also as a function of radius. The plots correspond to three cuts through the knot, namely cut~A (top panels), cut~B (middle panels) and cut~C (bottom panels). These are labelled in Figure~\ref{fig:circles}. 

Heating from atomic photoionization is dominant in the inner regions of the knot since the abundance of singly ionized helium (or neutral helium, in the case of cut~A) is highest there, due to the sixfold density increase in this region. The shadowing effect of the dense core is clearly visible in cut~A, causing heating by photoionization to decrease sharply already within the inner region of the knot. Heating by photoelectron emission from dust grain is higher in the envelope of the knot, where the radiation field is stronger. In fact, it is clear from Equation~\ref{eq:ggnu} that $G_{\rm d}(\nu)$, the frequency dependent contribution to the heating due to dust grains, is proportional to the flux, $F(\nu)$ and to the dust-to-gas ratio. As noted by the referee, the drop of 30\% in the value of the dust-to-gas ratio from the envelope to the core is not sufficient to explain the drop in the grain heating shown in Figure~\ref{fig:heat}, so this must be mainly due to the drop in $F(\nu)$, especially the attenuation of radiation at h$\nu~>~54.4$\,eV. In the shadowed region, $F(\nu)$ is dominated by the softer diffuse component of the radiation field, the stellar photons having been mostly absorbed within the dense core. For this reason dust heating will make a smaller contribution in the shadowed region, as confirmed by Figure~\ref{fig:heat}. 

It is worth re-iterating at this point that the emission from the surrounding nebular material is not accounted for in our model. This would also contribute to the diffuse radiation field impinging onto the shadowed region. However, due to the very low surface brightess of the nebular material, its contribution to the difffuse radiation field in the knot region is expected to be very small. 

\subsection{Dependence of the nebular models on the ionizing radiation spectrum}
\label{sub:differences}

\begin{table}
\caption{Comparison of results from photoionization models of knot J3 of Abell~30 with the dereddened observations using a non-LTE H-deficient model atmosphere (model~A) and a non-LTE hydrogen-rich atmosphere (model~B) as central ionizing star. The line intensities are given on a scale where I(He~{\sc i}~5876\,{\AA})\,=\,1. The {\it Observed} column lists the dereddened observed intensities, while the {\it A} and {\it B} columns list the results obtained, respectively, using H-deficient (model~A, as given already in Table~4) and H-rich (model~B) atmospheres for the central ionizing source.} \vspace{3mm}
\label{tab:dependence}
\begin{center}
\begin{tabular}{lccc}
\hline
Parameter		& Observed	& H-def		& H-rich 	\\
			& 		& star (A)	& star (B)	\\
\hline
He~{\sc i} 5876/10$^{-15}$(erg/cm$^2$/s)
			& 7.37		& 7.23		& 9.25		\\
He~{\sc i} 5876		& 1.		& 1.		& 1.		\\
He~{\sc ii} 4686 	& 2.50		& 2.11		& 2.12		\\
H~{\sc i} 4861		& 0.080		& 0.064		& 0.066		\\
C~{\sc ii} 4267		& 0.452		& 0.347		& 0.261		\\
C~{\sc iii}] 1908	& 2.8		& 1.52		& 0.958		\\
C~{\sc iii} 4648	& 0.030		& 0.046		& 0.025		\\
C~{\sc iii} 2297	& 0.960		& 1.53		& 0.687		\\
C~{\sc iv} 1550		& 9.2		& 15.4		& 8.02		\\
$[$N~{\sc ii}] 6584	& 0.829		& 0.141		& 0.390		\\
N~{\sc iv}] 1486	& 2.56		& 2.62		& 1.37		\\
$[$O~{\sc ii}] 3727	& 1.66		& 0.150		& 0.109		\\
O~{\sc ii} 4303		& 0.094		& 0.030		& 0.022		\\
O~{\sc ii} 4097		& 0.174		& 0.037		& 0.028		\\
O~{\sc ii} 4089		& 0.060		& 0.072		& 0.054		\\
O~{\sc ii} 4075		& 0.149		& 0.158		& 0.103		\\
$[$O~{\sc iii}] 5007	& 14.2		& 10.7		& 5.32		\\
$[$O~{\sc iii}] 4363	& 0.349		& 0.275		& 0.061		\\
$[$Ne~{\sc iii}] 3869	& 2.85		& 7.57		& 1.87		\\
$[$Ne~{\sc iv}] 2423	& 15.2 		& 18.22		& 7.28		\\
$[$Ne~{\sc v}] 3426	& 1.20		& 1.18		& 7.3		\\ 
\hline
\end{tabular}
\end{center}
\end{table}

As discussed in Section~2.2 we have used hydrogen-deficient synthetic spectra for the central star, calculated from line-blanketed non-LTE stellar atmospheres. The results of our nebular model (model~A) are compared, in Table~8, to those obtained using a H-rich stellar atmosphere model with the same values of T$_{eff}$ and log~g (model~B). In an attempt to isolate the effects that using a different ionizing spectrum for the central source might have on the nebular emission line spectrum, we have kept all other nebular parameters fixed (see Tables~1 and 2). The H-rich (H-Ni) atmosphere used for model~B is one of a set of model calculated with line blanketed NLTE model atmospheres \citep{werner86, rauch97}. The nebular emission line fluxes are listed, with the dereddened observed fluxes being given in the second column and the model values in the following two columns. The fluxes are given relative to He~{\sc i}~5876~{\AA} (which is given in units of 10$^{-15}$ ergs\,cm$^{-2}$\,s$^{-1}$ in row 1 of Table~\ref{tab:dependence}), on a scale where $I$(He~{\sc i}~5876~{\AA})\,=\,1. It is clear that the observed nebular spectrum is more closely reproduced by the photoionization model which used a hydrogen-deficient model atmosphere for the central star (model~A). The crucial failure of model~B is in the predicted  [O~{\sc iii}]~5007/4363 ratio of 87.2, versus model~A's value of 38.9, which more closely reproduces the observed value of 40.7. The much higher ratio predicted by model~B is indicative of the lower electron temperature obtained for the O~{\sc iii} region. 

\section{Conclusions}

We have constructed a photoionization model for the hydrogen-deficient knot J3 of Abell~30. This consisted of a spherical blob with a very dense metal rich core, surrounded by a layer of less dense gas, with somewhat less enhanced metal abundances. The envelope region composition might be formed by the mixing of the outer layers of the knot with the surrounding nebular gas. Our models found that chemical and density homogeneities alone were not  sufficient to create enough heating to explain the collisionally excited lines (CELs) observed in the spectrum of the knot \citep[see, for instance,][]{wesson03}. 

Following the hypothesis of \citet{borkowski91}, heating by photoelectric emission from dust grains was introduced into the model. Although using a simplified approach, this work has shown that for a grain radius of 10$^{-6}$\,cm a dust-to-gas ratio, $\rho_d/\rho_g$, of 0.107 by mass is sufficient to provide enough heating to the envelope of the knot to fit the CEL spectrum. Although the final value used for our model is about 10 times larger than the value of 0.01\,-\,0.02, thought to be typical for the interstellar medium, this is not surprising, as the actual amount of dust mixed with the gas in these types of object is very uncertain. \citet{borkowski91} inferred even larger dust-to-gas mass ratios, between 0.17 and 0.62, from their study of the dusty H-deficient planetary nebula in M22. However, as discussed in Section~2.2, the dust-to-gas ratio in the equatorial knots of Abell~30 is expected to be higher than in the polar knots \citep{harrington97}, hence higher than the value of 0.107 found here for the envelope the polar knot J3. 

Our final abundances derived for the two regions of the knot show core/envelope abundance ratios ranging from 5 for neon to 16 for carbon; these are somewhat less dramatic than the ratios of 300-700 found from the ORL/CEL abundance analysis of \citet{wesson03}, but nevertheless follow the same trend. It should also be noticed, when comparing our abundances with those derived by \citet{wesson03}, that the empirical results they obtained were based on the assumption of a chemically/density homogeneous knot, with a constant ionization structure. However, this is not likely to be the case since it is clear from Table~7 that the ionization structure is very different in the inner and outer regions of the knot. In particular, the abundance of oxygen derived from empirical analysis is very susceptible to the empirical ionization correction used and would change considerably if a bi-phase empirical analysis were to be carried out. Nevertheless, in agreement with Wesson et al.'s empirical analysis, our models find both phases of the knot to be oxygen-rich, with C/O ratios of 0.86 and 0.36 in the core and the envelope of the knot, respectively. This is in contrast to theoretical predictions for PNe in the born-again scenario which predicts C/O ratios larger than unity \citep{iben83, herwig01}.

\vspace{7mm}
\noindent

{\bf Acknowledgments}
The authors thank the referee for helpful comments on the relative gas and grain heating contributions. We also thank Mr R. Wesson for his help and advice. BE acknowledges support from PPARC Grant PPA/G/S/1997/00728 and the award of a University of London Jubber Studentship. TR acknowledges support from DLR grant 50\,OR\,0201.
\bibliographystyle{mn2e}

\bibliography{references}

\end{document}